\documentclass[12pt,nohyper]{article}         
\usepackage{amssymb,amsfonts}
\usepackage{epsfig}
\newcommand{\be}{\begin{equation}}
\newcommand{\ee}{\end{equation}}
\newcommand{\bea}{\begin{eqnarray}}
\newcommand{\eea}{\end{eqnarray}}
\newcommand{\bref}[1]{(\ref{#1})}
\newcommand{\nn}{\nonumber}
  
 \newcommand{\D}{\delta} 
\newcommand{\ep}{\epsilon}

         \newcommand{\lam}{\lambda}
           \newcommand{\s}{\sigma}
          \newcommand{\w}{\omega}

\newcommand{\W}{\Omega}

\newcommand{\ba}{\overline }
\def\6{\partial}
\def\7{\tilde}\def\t{\widetilde}
\def\8{\widehat}
 
\def\vs{\vskip 3mm}
\def\eom{EL }
\newcommand{\M}{M}
\def\pa{\partial}

\def\CD{{\cal D}}\def\CL{{\cal L}}
\def\CE{{\cal E}}

\def\l{{\ell}}
\def\lag{Lagrangian }\def\lags{Lagrangians }
\def\ham{Hamiltonian }

\begin{document}


\begin{center}{\Large \textbf{Physical Degrees of Freedom of Non-local Theories}}\end{center}

\vskip 4mm

\begin{center}{\Large Joaquim Gomis$^a$, Kiyoshi Kamimura$^b$, Toni  Ram\'{\i}rez$^a$}\end{center} 

\vskip 4mm

\begin{center}${}^a${Departament ECM, Facultat de F{\'\i}sica, \\

Universitat de Barcelona, Institut de F{\'\i}sica d'Altes Energies, \\

and CER for Astrophysics, Particle Physics and Cosmology, \\

Diagonal 647, E-08028 Barcelona, Spain }\\

\end{center} \begin{center}

${}^b${Department of Physics, Toho University, 

        Funabashi\ 274-8510, Japan\\

\medskip 

{\small E-mail: gomis@ecm.ub.es, kamimura@ph.sci.toho-u.ac.jp, 

tonir@ecm.ub.es }} \end{center}

\begin{abstract}We analyze the physical reduced space of non-local theories,
around the fixed points of these systems, by analyzing:
i) the Hamiltonian constraints appearing in the 1+1 formulation,  
ii) the symplectic two form in the surface on constraints.

P-adic string theory for spatially homogeneous configurations has
two fixed points.
The physical phase space around $q=0$ is trivial, instead around
 $q=\frac 1g$ is infinite dimensional. For the special case of the rolling
tachyon solutions it is an infinite dimensional lagrangian submanifold.
In the case of string field theory, at lowest truncation level, the
physical phase space of  spatially homogeneous configurations
is two dimensional around $q=0$,
which is the relevant case for the rolling tachyon solutions,
and infinite dimensional around $q=\frac {M^2}g$.
\end{abstract}


\section{Introduction}\label{sec.1}
\noindent

Recently it has been a lot of interest to study open string tachyon
condensation. There are
several approaches to study this problem like
string field theory, p-adic string theory, boundary conformal field theory,
Born-Infeld effective theory, non-critical string theory, matrix models,
see for example
\cite{Sen:1999nx},\cite{Ghoshal:2000dd},\cite{Sen:1999md},\cite{Moeller:2002vx},\cite{Lambert:2003zr},\cite{fujitahata:2003fh},\cite{McGreevy:2003kb},\cite{Klebanov:2003km},\cite{Schomerus:2003vv},\cite{Karczmarek:2003xm}.
In ref \cite{Moeller:2002vx}  the
rolling of the tachyon has been analyzed by constructing a
classical time dependent
spatially homogeneous solution of p-adic string theory (p-adic particle).
This solution has oscillations with {time} and
ever-growing amplitude $\phi(t)=\sum_n a_n e^{ nt}$.

Since the string field theory \cite{Witten:1985cc} and the
p-adic string theory \cite{Brekke:dg} are non-local
theories
the construction of solutions  and the initial value
problem, which is related to the dimension of the
physical reduced phase space, are non-trivial 
issues.\footnote{
A non-singular \lag system up to $n$-th time derivatives has
$2n$ degrees of freedom. The canonical description was given by
Ostrogradski \cite{o}. In general non-local systems
have infinite degrees of freedom and their energies are not bounded.
Naively in order to get solutions,
in a generic point, one needs an infinite number of initial conditions. }
{ Some aspects of non-local theories in connection with the string 
filed theory have been examined }\cite{Eliezer:1989cr}.

In this paper, motivated by these problems, we present a
general method to analyze reduced phase space of non-local theories.
It is based on the
1+1 dimensional Hamiltonian formalism of non-local theories
{proposed in \cite{lv}, and further developed in \cite{Gomis:2000gy}.}
The formalism consists of a two dimensional field theory
Hamiltonian and two sets  of phase space
constraints, momentum  and Euler Lagrange (EL) constraints.
In this framework the Euler Lagrange equation of motion
appears as a Hamiltonian constraint.
The reduced phase space is constructed by analyzing
the first or second class character of these constraints.
The analysis is analogous to the construction of the physical
phase space of gauge theories.
 As we will see there are two types of reduced phase spaces
around any fixed point
that we call perturbative and non-perturbative.
In the first case we solve the second class
constraints in a smooth way for $g\rightarrow 0$, where $g$ is
a  coupling constant of the non-local theory,
and in the second case we solve non-perturbatively
the second class constraints.
An alternative way to analyze the structure of the
reduced phase space consists in computing the
infinite dimensional symplectic two form on the surface defined by the
momentum and EL constraints, around the fixed points.

\begin{figure}[ht]
    \includegraphics[width=90mm,height=50mm,clip]{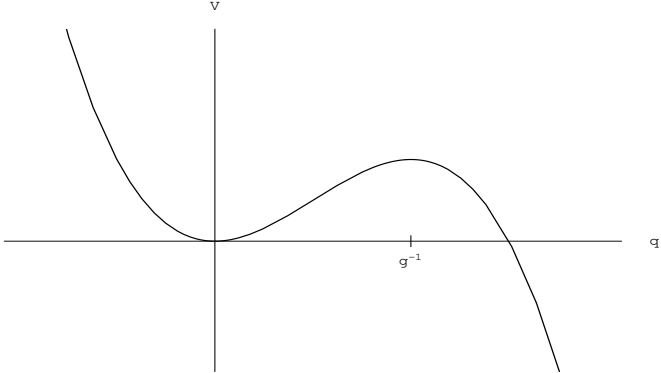}
\caption{{\it Potential of P-adic Theory.}
The phase space is zero dimensional at the minimum $q=0$.
It is infinite dimensional at the local maximum $q=1/g$.
For the rolling solutions it is infinite dimensional lagrangian submanifold.
}

\end{figure}
\begin{figure}[ht]
\includegraphics[width=90mm,height=50mm,clip]{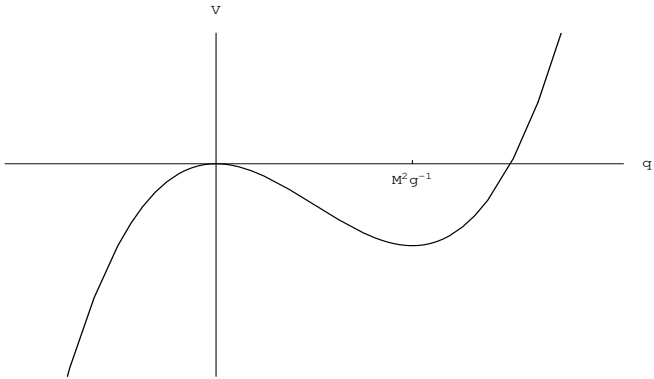}
\caption{{\it Potential of String Field Theory.}
The phase space is two dimensional at the local maximum $q=0$.
It is infinite dimensional at the local minimum $q=M^2/g$.
}\end{figure}

In the case of smooth
spatially homogeneous configurations of p-adic string theory, we will see
that the perturbative physical reduced phase space around the trivial
fixed point $q=0$ is zero dimensional,
where all the coordinates and momenta vanish (fig.1).
There is no mode excitation around this minimum.
For the fixed point $q=\frac 1g$
the perturbative reduced phase space is infinite dimensional.
The unstable and stable manifold of this non-trivial fixed point
is a lagrangian submanifold of infinite dimensions, where the
symplectic form vanishes. This implies that the manifold of solutions
relevant for the rolling tachyon, unstable submanifold, is infinite
dimensional without a phase space structure. 
If we do not impose the boundary condition, solutions do not have  
a definite sign of the energy. This is a general property  for a generic situation in non-local theories.

We will also analyze the zero-th level truncation of
string field theory as a non-local theory\cite{Eliezer:1989cr}.
In this case the perturbative reduced phase space around
$q=0$ is two dimensional (fig.2).
This implies that the space of solutions relevant for the rolling
tachyon is two dimensional. 
The perturbative phase space around the fixed point $q=\frac {M^2}g$
is infinite dimensional. This implies that there are infinite excitation
modes around this minimum.
The unstable and stable manifold of this non-trivial fixed point
is a lagrangian submanifold of infinite dimension where the
symplectic form vanishes.
The difference with respect to the p-adic case is due to the different
dispersion relation for the solutions we found.

The non-local harmonic oscillator is examined to illustrate the formalism. 
It have been studied in detail in \cite{Eliezer:1989cr} and \cite{lv}
and is known to be a finite dimensional system.
Around the only one fixed point $q=0$ we
reproduce that the phase space is
two dimensional perturbatively and four dimensional non-perturbatively.
In these cases we will compute the energy for the set of  solutions
verifying the EL equations.

In section 2 a brief review of the 1+1 dimensional Hamiltonian
formalism for non-local theories is introduced. In section 3
we apply it to  a non-local harmonic oscillator and show how the
infinite dimensional phase space is reduced using the constraints.
In sections 4 and 5 we discuss the p-adic and string field theories.
Discussions will be in the last section.
\vs


\section{1+1 Dimensional Formalism of Non-local Theories}
\noindent

Ordinary local \lags depend on a finite number of derivatives at a given time,
it is
\be
{L}(q(t),\dot{q}(t),\cdots,q^{(n)}(t)).
\ee
Now, we can consider a \lag depending on a piece of the trajectory
$q(t+\lambda), \;\;\; \forall \lambda$ belonging to an interval $[ a,b ]$,
where $a$ and $b$ are real numbers

\be
{L}(q(t+\lambda)).
\ee
This is how we obtain a non-local \lag,
that we denoted by $L^{non}(t)$.

For sufficiently
smooth trajectories we can expand $q(t+\lambda)$ to all powers in
$\lambda$. In this case, this \lag
can be written as a function of all time derivatives $\frac{d^m}{dt^m}q(t)$
with $m=0,1,2,\cdots$.
The EL equation is
\be \label{ELeq}
\int dt {\D L^{non}(t) \over \D q(t')} = 0.
\ee
This equation should be understood as a functional relation to be satisfied by
physical trajectories, i.e., a \lag constraint.
It defines a subspace $J_R$
of physical trajectories
\be
J_R \subset J
\ee
in the space of all possible trajectories $J$.
There is no dynamics except the displacement inside the trajectory
\be
q(t) \stackrel{T_\lambda}{\rightarrow} q(t+\lambda).
\ee

The $1+1$ dimensional formalism of non-local \lags
\cite{lv},\cite{Gomis:2000gy} is based in considering a field $Q(t,\lambda)$,
instead of trajectory $ q(t)$, with the restriction \footnote{
{An equivalent formalism was also given in \cite{Woodard:2000bt} for a nonlocal
system of finite extend.} }

\be
\dot Q(t,\lambda)=Q'(t,\lambda)
\label{QQeq}
\ee
where $\dot Q(t,\lambda)=\frac{\pa}{\pa t}Q(t,\lambda)$ and $
Q'(t,\lambda)=\frac{\pa}{\pa \lambda}Q(t,\lambda)$.
Equation \bref{QQeq}
implies $Q(t,\lambda)=F(t+\lambda)$, we further assume $F(t)=q(t)$.
The  Hamiltonian is given by
\be
H(t,[Q,P])=\int d\lam P(t,\lam)Q^\prime(t,\lam)- {\cal L}(t,0),
\label{h}
\ee
where $P(t,\lam)$ is the canonical momentum of
$Q(t,\lam)$. The phase space is thus $T^*J$
with the Poisson brackets
\be
\{Q(t,\lambda),  P(t,\lam')\}=\delta (\lambda-\lambda').
\label{pb1}
\ee
The symplectic two form $\Omega$ is given by
\be
\Omega=\int d\lam~ dP(t,\lam)\wedge dQ(t,\lam).
\label{twoform}
\ee
The ``\lag density" $\CL(t,\lam)$ is constructed from the original
non-local \lag $L^{\rm non}(t)$
by replacing
$q(t)$ by $Q(t,\lam)$, $t$-derivatives of $q(t)$ by $\lam$-derivatives
of $Q(t,\lam)$ and $q(t+\rho)$ by $Q(t,\lambda+\rho)$.

There are two sets of constraints in the formalism.
One is associated with the definition of the momentum $P(t,\lambda)$,
which for a non-local theory becomes a Hamiltonian constraint
(momentum constraint)
\be
\varphi(t,\lam,[Q,P]) \equiv P(t,\lam) -
\int \,d\sigma \, \chi(\lam,-\sigma) \,\CE(t;\s,\lam) ~\approx 0 \,,
\label{mom}
\ee
where $\CE(t;\s,\lam)$ and $ \chi(\lam,-\s)$ are defined by
\bea
\CE(t;\s,\lam)&=&\frac{\delta \CL(t,\s)}{\delta Q(t,\lam)},~~~
\chi(\lam,-\s)~=~\frac{\ep(\lam)-\ep(\s)}{2}.
\eea
The other is equivalent to the Euler Lagrange equation, (\eom constraint),
 explicitly
\be
\psi(t,\lam,[Q]) \equiv \int d\s~\CE(t;\s,\lam) \approx 0 \,.
\label{eom}
\ee
In fact if we use \bref{QQeq}, $\psi\approx 0$ implies the EL equation of the
original non-local \lag $L^{\rm non}(t)$.

The momentum constraints \bref{mom} are stable by using the \eom constraints 
\bref{eom}.
The stability
of the \eom constraints is guaranteed by itself,
\bea
\dot\varphi~=~\{\varphi,H\}&\sim&\psi~\approx~0,~~~~
\dot\psi~=~\{\psi,H\}~\sim~\psi~\approx~0.
\eea
Therefore the Hamiltonian \bref{h}, the momentum constraint
\bref{mom} and the \eom constraint \bref{eom} define a consistent
 Hamiltonian system \cite{Dirac}.

It is local with respect to the "time" $t$ although it
appears non-locality with respect to the "space" $\lam$.
The momentum constraint
\bref{mom} and the \eom constraint \bref{eom} define a surface $\Sigma^*$
in the $T^*J$.

\vs
In order to analyze the reduced phase space for smooth trajectories,
it is convenient to make an expansion of the coordinate and
momentum as
\bea
Q(t,\lam)&=&\sum_{m=0}^\infty~e_m(\lam)~q^m(t),~~~~~~
P(t,\lam)~=~\sum_{m=0}^\infty~e^m(\lam)~p_m(t),
\label{QPexp}\eea
where $e_m(\lam)$ and $e^m(\lam)$ are ortho-normal basis
\bea
e_m(\lam)~=~\frac{\lam^m}{m!},~~~&&~~~
e^m(\lam)~=~(-\pa_\lam)^m\delta(\lam),
\label{expbase}\\
\int d\lam~e^m(\lam)~e_\l(\lam)~=~{\D^m}_\l,~~~&&~~~
\sum_{m=0}^{\infty}~e^m(\lam)~e_m(\lam')~=~\D(\lam-\lam').
\label{ortho}
\eea
The Poisson brackets for the variables $q^m,p_n$ are given by
\be
\{ q^m, p_n \}= {\delta^m}_n.
\label{pb2}
\ee
The symplectic two form \bref{twoform} is given by
\be
\Omega=\sum_{m=0}^\infty dp_m\wedge dq^m.
\label{twoform1}
\ee

The Hamiltonian \bref{h} can be written as
\be
H(q,p)=\sum_{m=0}^\infty p_mq^{m+1}-L(q^0, q^1,...).
\label{hexp}\ee
One of the Hamilton's equation implies $ q^{m+1}=\frac d{dt} q^m$.
The momentum constraint $\varphi(t,\lam)$
can be expanded in terms of $e^m(\lam)$ as
\be
\varphi(t,\lam)= \sum_{m=0}^\infty~\varphi_m(t)e^m(\lam)
\label{momexp}\ee
and the \eom constraint $\psi(t,\lam)$ can be expanded by $e_m(\lam)$ as
\be
\psi(t,\lam)= \sum_{m=0}^\infty~\psi^m(t)e_m(\lam).
\label{eomexp}\ee
Our formalism applied to
local \lags depending on finite number of time derivatives
reproduces the Ostrogradski formalism \cite{o}.

\subsection{Reduced phase space}
\indent

In order to examine the reduced phase space
and find the physical degrees of freedom,
it is crucial to analyze the first and second class character of the
constraints,
the rank of the matrix of the Poisson brackets
\bea
\pmatrix{\{\varphi(\lam),\varphi(\lam')\},&\{\varphi(\lam),\psi(\lam')\}\cr
\{\psi(\lam),\varphi(\lam')\},&\{\psi(\lam),\psi(\lam')\}\cr},
\eea
or equivalently  in terms of $ \varphi_m$ \bref{momexp}
and $\psi^m $ \bref{eomexp},
\bea
\pmatrix{\{\varphi_m,\varphi_\l\},&\{\varphi_m,\psi^\l\}\cr
\{\psi^m,\varphi_\l \},&\{\psi^m,\psi^\l\}\cr},
\eea
around the trivial and non-trivial fixed points.
 The unphysical degrees of freedom
are eliminated by solving the second class constraints and using the
associated Dirac brackets.
We can also write the Hamiltonian in the reduced phase space.

This procedure is analogous to the construction of the physical phase space
of gauge theories. In this case one introduces gauge fixing constraints
in such a way
to convert the first class constraints, associated to gauge invariances,
to second class constraints. The gauge degrees of freedom are eliminated by
solving the second class constraints.

When  a non-local theory under investigation contains an infinite number
of second class constraints, the dimension of the physical phase space
 might be
reduced to be finite dimensional. This implies that with
a suitable choice of variables
the system  would be described in terms of a local theory.

An alternative way to see the physical degrees of freedom
is to compute the symplectic form, \bref{twoform} or \bref{twoform1},
on the surface $\Sigma^*$ defined by the constraints
\bref{momexp} and \bref{eomexp}.
\vs


\section{Non-local harmonic oscillator}
\noindent

{In this section we will apply the above formalism to an harmonic oscillator 
with a non-local interaction to illustrate
how the reduced phase spaces come out within this frameworks.
We will compute 
the perturbative and the non-perturbative phase space for this model
around the unique fixed point $q=0$.
The dimension of the reduced phase space is 
finite dimensional in agreement with }\cite{Eliezer:1989cr},\cite{lv}.

\vs
The \lag of the non-local harmonic oscillator is given by

\bea
L^{non}(q(t))&=&\frac12\dot q(t)^2~-~\frac{\w^2}2 q(t)^2~+~
\frac{g}{4}\int dt'q(t)e^{-|t-t'|}q(t').
\label{har}
\eea
The associated 1+1  dimensional \lag density $\CL(t,\lam)$ is given by
\bea
\CL(t,\lam)&=&\frac1{2} Q'(t,\lam)^2~-~\frac{\w^2}2 Q(t,\lam)^2~+~\frac{g}{4}
\int d\lam'Q(t,\lam)e^{-|\lam-\lam'|}Q(t,\lam')
\eea
and the Hamiltonian \bref{h} is
\bea
H&=&
\sum_{m=0}^\infty  p_m  q^{m+1}~-~
[\frac{1}2 (  q^1)^2~-~\frac{\w^2}2 (  q^0)^2~+~\frac{  g }{2}~
q^0\sum_{r=0}{q^{2r}}~].
\label{hhoexp}
\eea
The momentum constraint \bref{mom} in the basis \bref{momexp} implies
the constraints
\bea
 \varphi_{2m}&=&  p_{2m}~-~ q^1~\D_{0,m}~+~\frac{  g}{2}\sum_{r=0}^\infty~
q^{2r+1} ~\approx~0
,
\nn\\
 \varphi_{2m+1}&=&  p_{2m+1}~-~\frac{  g}{2}\sum_{r=0}^\infty~  q^{2r}~
\approx~0,~~~~~~~~~~~~~~~(m\geq 0).
\label{momhoexp}
\eea
The \eom constraint \bref{eom} is
\bea
 \psi(t,\lam)&=&Q''(t,\lam)~+~{\w^2}Q(t,\lam)~-~
\frac{g}{2}\int d\lam'~e^{-|\lam-\lam'|}Q(t,\lam')~\approx~0,
\label{eomho}
\eea
in the basis \bref{eomexp} it implies
\bea
 \psi^m&=&q^{m+2}~+~{\w^2}  q^m~-~{  g}
\sum_{r=0}^\infty~{  q}^{m+2r}~\approx~0,~~~~~~~~(m\geq 0).
\label{eomhoexp}
\eea
Note $\frac d{dt}\psi^m=\psi^{m+1}$, i.e the constraints $\psi^m$ preserve
their form in time.
The \eom constraint \bref{eomho} with the Hamilton's equation, \bref{QQeq},
$\dot Q=Q'$, i.e.,
$ q(t+\lam)\equiv Q(t,\lam)$
reproduces the \eom equation of the non-local action \bref{har},
\bea
\ddot q(t)~+~{\w^2}q(t)~-~
\frac{g}{2}\int dt'~e^{-|t-t'|}q(t')&=&0.
\label{eomnon}
\eea
\vs

\subsection{Simple Harmonic Oscillator}
\noindent

When the coupling constant $g$ vanishes this system is an ordinary harmonic
oscillator. The constraints \bref{momhoexp} and \bref{eomhoexp} are
\bea
\varphi_0&=&p_0-q^1~\approx~0,~~~~~~~~~\varphi_m~=~p_m~\approx~0,~(m\geq 1),
\\
\psi^m&=&q^{m+2}+\w^2 q^m~\approx~0,~~~~~~(m\geq 0).
\label{simple}\eea
They are second class constraints and are paired as
\bea
\varphi_0&=&p_0-q^1,~~~{\rm with}~~~\varphi_1~=~p_1
\label{scfho}\eea
to eliminate the canonical pair $(q^1,p_1)$.
Analogously,
\bea
\varphi_{m+2}&=&p_{m+2},~~~{\rm and}~~~\psi^m~=~q^{m+2}+\w^2 q^m
\label{scfho2}
\eea
are paired to eliminate the canonical pairs  $(q^{m+2},p_{m+2})$ for
${m+2}\geq 2$. As a result, all canonical pairs  $(q^{m},p_{m})$ with
${m}\geq 1$ are expressed
in terms of the canonical pair $(q^0,p_0)$.
The Dirac bracket in the  reduced phase space of $(q^0,p_0)$
coincides with the Poisson bracket,
since the sets of the second class
constraints \bref{scfho} and \bref{scfho2} have the standard
forms{\footnote{
When a set of second class constraints has the form
$\{p=0,q=f({\bf q},{\bf p})\} $ the canonical pair $(p,q)$ is expressed
in terms of the remaining variables $({\bf p},{\bf q})$.
The  Dirac bracket is defined by
$$\{A,B\}^*= \{A,B\}-\{A,p\}\{q-f({\bf q},{\bf p}),B\}+
\{A,q-f({\bf q},{\bf p})\}\{p,B\}.$$
The values of Dirac bracket of $({\bf q},{\bf p})$ are same as those of the
Poisson bracket.
We refer them as a standard form of second
class constraints.
 $\{q=0,p=g({\bf q},{\bf p})\}$ is also
the standard form of second class constraints.}.
The Hamiltonian $H^*$ in the reduced space $\Sigma^*$ becomes the one of an
ordinary harmonic oscillator
\bea
H^*&=&
\frac12(p_0)^2~+~\frac{\w^2}2 (q^0)^2.
\label{hfho}
\eea
The dimension of the reduced phase space is $2$ ( one canonical pair )
as was expected.
 \vs

\subsection{Perturbative Pairing}
\noindent

When the coupling constant $g$ is small and the perturbative treatment is
allowed the physical degrees of freedom of the system are same as in the
$g=0$ case. This is because
all constraints remain  in the second class and we can make
the same pairing of
constraints as in the free case.
To show it  explicitly we should rewrite the constraints
in the standard form.

The \eom constraints \bref{eomhoexp} are expressed using by themselves
iteratively as
\bea
\t\psi^{2\l}&=&q^{2\l+2}+(-1)^\l k^{2\l+2} q^0~\approx~0,~~~~~~
\nn\\
\t\psi^{2\l+1}&=&q^{2\l+3}+(-1)^\l k^{2\l+2} q^1~\approx~0,
\label{eomho2}
\eea
where
\bea
k^2&=&
\w^2-\frac{g}{(\w^2+1)}+\frac{g^2}{(\w^2+1)^3}
-\frac{2~g^3}{(\w^2+1)^5}+...~=~
\frac{(\w^2-1)+\sqrt{(\w^2+1)^2-4g}}{2}.
\nn\\ \label{solk}
\eea
The two $k$'s given from \bref{solk} are solutions of the
characteristic equation associated with the EL equation, \bref{eomnon},
\bea
k^2 -\w^2+g~\frac{1}{1+k^2}~=~0.
 \label{dispho}\eea
The other two solutions of \bref{dispho} can not be obtained by
iteration in $g$.
(See next subsection.)
Instead of working with the momentum constraints \bref{momhoexp} it is
more convenient to introduce the following combinations
\bea
\t\varphi_{2\l}&=&
\varphi_{2\l}-\frac{g}{2}\sum_{r=0}^\infty~\t\psi^{2r+1}~=~
  p_{2\l}-q^1~\D_{\l,0}+
\frac{g}{2}\frac{q^1}{1+k^2}~\approx~0,~~~~
\nn\\
 \t\varphi_{2\l+1}&=&
\varphi_{2\l+1}+\frac{g}{2}\sum_{r=0}^\infty~\t\psi^{2r}~=~
p_{2\l+1}-\frac{  g}{2}\frac{q^0}{1+k^2}~\approx~0 .
\label{momho2}
\eea
In order to rewrite the constraints in the standard form we perform
a canonical transformation generated by
\bea
W&=&\sum_{m=0}^\infty
\left(q^{2m}(\t p_{2m}-\frac{g}{2}\frac{q^1}{1+k^2})+
q^{2m+1}(\t p_{2m+1}+\frac{g}{2}\frac{q^0}{1+k^2})\right)+
\frac{g}{2}\frac{q^0q^1}{(1+k^2)^2}
\nn\\ \eea
which gives
\bea
\t q^\l&=&q^\l,
\nn\\
p_{0}&=&\t p_{0}-\frac{g}{2}\frac{q^1}{1+k^2}+{g}\frac{q^1}{(1+k^2)^2}
+\frac{g}{2{(1+k^2)}}\sum_{r=0}^\infty~\t\psi^{2r+1}
,
\nn\\~~~~
p_{1}&=&\t p_{1}+\frac{g}{2}\frac{q^0}{1+k^2}-
\frac{g}{2{(1+k^2)}}\sum_{r=0}^\infty~\t\psi^{2r}
,
\nn\\
p_{2m}&=&\t p_{2m}-\frac{g}{2}\frac{q^1}{1+k^2},~~~~
p_{2m+1}=\t p_{2m+1}+\frac{g}{2}\frac{q^0}{1+k^2},~~~(m\geq 1).
\nn\\
\eea
In terms of the new canonical variables we have
\bea
\t\varphi_0&=&\t p_0-(1-\frac{g}{(1+k^2)^2})\t q^1~\approx~0,~~~~~~
\t\varphi_1~=~\t p_1~\approx~0,~~~~~~
\nn\\
\t\varphi_{2\l+2}&=&\t p_{2\l+2}~\approx~0,~~~~~~
\t\psi^{2\l}~=~q^{2\l+2}+(-1)^\l k^{2\l+2} \t q^0~\approx~0,
\nn\\
\t\varphi_{2\l+3}&=&\t p_{2\l+3}~\approx~0,~~~~~~
\t\psi^{2\l+1}~=~q^{2\l+3}+(-1)^\l k^{2\l+2}\t q^1~\approx~0,~~~(\l\geq 0).
\label{cc}
\eea
The constraints \bref{cc}
 are used to eliminate canonical pairs $(\t q^m,\t p_m),~(m \geq 1)$
in terms of $(\t q^0,\t p_0)$. The Dirac bracket between $(\t q^0,\t p_0)$ is
$\{\t q^0,\t p_0\}^*=1$. Thus the reduced phase space has dimension $2$
and the Hamiltonian
is
\bea
H&=&\frac12\left(\frac{(1+k^2)^2}{(1+k^2)^2-g}~(\t p_0)^2~+~
\frac{(1+k^2)^2\w^2-g(1+2k^2)}{(1+k^2)^2}({\t q}^0)^2\right).
\eea
In the reduced space we have an ordinary
 harmonic oscillator with the frequency $k$,
\bea
\frac{(1+k^2)^2}{(1+k^2)^2-g}~\frac{(1+k^2)^2\w^2-g(1+2k^2)}{(1+k^2)^2}
&=&k^2,
\eea
and mass
\be
M=\frac{(1+k^2)^2-g} {(1+k^2)^2}.
\ee

\vs


\subsection{Non Perturbative Pairing}
\noindent

If the coupling constant is not small there is another possible pairing
for the constraints \bref{momhoexp} and \bref{eomhoexp}.
Let us consider a canonical transformation generated by
\bea
W&=&\t p_0q^0+\t p_1(\sum_{j=0}q^{2j+1})+
                   \t p_2(\sum_{j=0} q^{2j+2})+\sum_{j=3}\t p_jq^j-
\frac{g}{2}(\sum_{j=0}q^{2j+1})(\sum_{\l=0} q^{2\l+2}).
\nn\\
\eea
It makes $(\sum_{j=1,odd}q^j)$ and $(\sum_{j=2,even}q^j)$ to be new
coordinates,
\bea \matrix{
\t q^0&=&q^0,~~~&p_0&=&\t p_0 \cr
\t q^1&=&(\sum_{j=1,odd}q^j),&p_1&=&\t p_1 -\frac{g}{2}\t q^2\cr
\t q^2&=&(\sum_{j=2,even}q^j),&p_2&=&\t  p_2 -\frac{g}{2}\t q^1\cr
\t q^{2\l+1}&=&q^{2\l+1},~~~ &p_{2\l+1}&=&\t p_{2\l+1}+\t p_1 -
\frac{g}{2}\t q^2,&~
~~~(\l\geq 1) \cr
\t q^{2\l}&=&q^{2\l},~~~ &p_{2\l}&=&\t p_{2\l}+\t p_2-\frac{g}{2}\t q^1,
&~~~~~~(\l\geq 2).
}
\eea
In terms of new variables the constraints \bref{momhoexp} and \bref{eomhoexp}
are
\bea
\varphi_0&=&\t p_0~-~(\t q^1-\sum_{j=1}\t q^{2j+1})+\frac{g}{2}\t  q^1,
\nn\\
\varphi_1&=& \t p_1~-\frac{g}{2}\t q^0 -{g}\t  q^2, ~
\nn\\
\varphi_{\l}&=&\t p_{\l},~~~~~~~~~(\l\geq 2).
\label{nhomom}
\eea
\bea
\psi^0&=&(1-g)\t q^{2}+(\w^2-g)\t q^{0}-\sum_{j=2}\t q^{2j},
\nn\\
 \psi^1&=&(1-{\w^2})\t q^{3}~+~({\w^2}-g)\t q^1~-~{\w^2}\sum_{j=2}\t q^{2j+1}
,
\nn\\
 \psi^2&=&(1-\w^2)\t q^{4}~+~(\w^2-g)\t q^{2}~-~\w^2\sum_{j=2}\t q^{2j+2},
\nn\\
 \psi^\l&=&(1-g)\t q^{\l+2}~+~({\w^2}-g)  \t q^\l~-~{  g}\sum_{r=2}~
{\t  q}^{\l+2r},~~~~~~(\l\geq 3).
\label{EOMconst}
\eea
The Hamiltonian \bref{hhoexp} becomes
\bea
H&=&\t p_0(\t q^1-\sum_{r=1}\t q^{2 r+1})
+(\t p_1-\frac{g}{2}\t q^2) \t q^2+(\t p_2-\frac{g}{2}\t q^1)
\sum_{r=1}\t q^{2 r+1}
+\sum_{r=3,odd} \t p_r  \t q^{r+1}
\nn\\&+&\sum_{r=4,even} \t p_r  \t q^{r+1}
-
[\frac{1}2 (\t q^1-\sum_{r=1}\t q^{2 r+1})^2~-~\frac{\w^2}2 (\t q^0)^2~+~
\frac{  g }{2}~\t q^0~(\t q^0+\t q^2)].
\label{nhohamilt}
\eea
\vs

There are two inequivalent pairings:
 \bea
i),&&~~~~~(\varphi_0,\varphi_1),~~~(\varphi_{\l+2},\psi^\l),~~(\l\geq 0),
\eea
it is the perturbative pairing discussed in the  previous subsection.
\bea
ii),&&~~~~(\varphi_0,\varphi_3),~~~(\varphi_1,\varphi_2),~~~
(\varphi_{\l+4},\psi^\l),~~(\l\geq 0).
\eea
It is a non-perturbative paring because it requires an inverse power of $g$.
$\varphi_1$ and $\varphi_2$ are paired to eliminate $(\t  q^2,\t  p_2)$ as
\bea
\t p_2&=&0,~~~~~~\t q^2~=~\frac{1}{g}~\t p_1-\frac12 \t q^0.
\label{p2q2}
\eea
$\varphi_0$ and $\varphi_3$ are paired to eliminate $(\t q^3,\t p_3)$ as
\bea
\t p_3&=&0,~~~~~~\sum_{r=1}\t q^{2 r+1}~=~-\t p_0+ \t q^1-\frac{g}{2}\t q^1.
\label{elhnp}
\eea
Other constraints, $(\varphi_{\l+4},\psi^\l),~(\l\geq 0)$,
are used to solve $(\t q^j,\t p_j),(j\geq 4)$.
Thus the reduced phase space is spanned by  $(\t q^j,\t p_j),(j=0,1)$ and
is 4 dimensional.

The Hamiltonian in the reduced space is
\bea
H&=&\frac12(\t p_0+\frac{g}{2}\t q^{1})^2
+\frac{g}{2}(\frac{1}{g}\t p_1-\frac12 \t q^0)^2
-\frac{g}{2}(\t q^1)^2+\frac{\w^2}2 (\t q^0)^2~-~
\frac{  g }{2}(\t q^0)^2.
\label{NPHam}
\eea
The non-vanishing Dirac brackets for $ (\t q^0,\t q^1,\t p_0,\t p_1)$
are  given by $\{\t q^0, \t p_0\}^{*}=\{\t q^1, \t p_1\}^{*}=1$.
Further canonical transformation shows
it is a system of two harmonic oscillators
\bea
H&=&\frac12(\8 p_0)^2+\frac{k_+^2}{2}(\8 q^{0})^2~+~
\frac12(\8 p_1)^2+\frac{k_-^2}{2}(\8 q^{1})^2,~~~~
\nn\\&& k^2_\pm=\frac{(\w^2-1)\pm\sqrt{(\w^2+1)^2-4g}}{2}.
\label{NPHam2}
\eea
The frequencies $k^2_\pm$ are solutions of \bref{dispho}.
For small value of $g$, $k_+^2>0$ represents the ordinary harmonic oscillator
mode while  $k_-^2<0$ mode has potential of inverse sign.

Summing up, the reduced perturbative phase space around $q=0$ is
two dimensional,
which is the same dimension as the free theory $(g=0)$.
Instead the non-perturbative one is four dimensional.
Notice that in both cases there is an infinite dimensional reduction of the
original degrees of freedom.
\vs


\section{P-adic particle}
\noindent

P-adic string theory \cite{Brekke:dg}
 has been used as a toy model to study tachyon
condensation.
For spatially homogeneous configurations (p-adic particle) the
rolling of the tachyon has been analyzed by constructing a
classical time dependent solution .
This solution has oscillations in time with
ever-growing amplitude $\phi(t)=\sum_n a_n e^{ nt}$
\cite{Moeller:2002vx}. Here we will analyze
the reduced phase space of the p-adic particle.

The action  for the p-adic particle is
\bea
S&=& \frac{1}{g_p^2}~\int dt~[-\frac12~q(t)~p^{\frac12{\pa_{t}^2}}~q(t)~
+\frac1{p+1}~q(t)^{p+1}~],
\eea
where $\frac{1}{g_p}=\frac{1}{g^2}\frac{p^2}{p-1}$ and  $p$ is a prime number.
The Euler-Lagrange equation is
\be \label{paELeq}
p^{\frac{\pa_{t}^2}{2}}~q(t)~=q(t)^{p}.
\ee
\vs

In the following we examine the $p=2$  system.
After rescaling the \lag density, $\CL(t,\lam)$
can be defined by
\bea
\CL(t,\s)&=&-\frac12~Q(t,\s)~e^{\pa_\s^2}Q(t,\s)~+~
\frac{g}{3}~Q(t,\s)^3.
\label{pLag}\eea
The \eom constraint \bref{eom} is
\bea
\psi(t,\lam,[Q])\equiv \int d\s~\CE(t;\s,\lam) &=&
-~e^{\pa_\lam^2}~Q(t,\lam)~+~g~Q(t,\lam)^2~\approx~0
\label{peom}
\eea
and the momentum constraint \bref{mom} becomes
\bea
\varphi(t,\lam,[Q,P]) &\equiv& P(t,\lam) -
\int \,d\sigma \, \chi(\lam,-\sigma) \,\CE(t;\s,\lam)
\nn\\&=&
P(t,\lam) ~-\frac12~
\sum_{m=0}^{\infty}\sum_{\l=0}^{\infty}
\frac{1}{(\l+m+1)!}
[(\pa_\s^{2\l+1}Q(t,\s))|_{\s=0}~(-\pa_\lam)^{2m}\D(\lam)
\nn\\ &  &~~~~~~~~~~~~~~~~~~
-~(\pa_\s^{2\l}Q(t,\s))|_{\s=0}~(-\pa_\lam)^{2m+1}
\D(\lam)]
\approx 0,\label{pmom}
\eea
where we have used,
\bea
\chi(\lam,-\s)\pa^\l_\s\D(\lam-\s)&=&\sum_{m=0}^{\l-1}(-)^m\pa_\s^
{\l-1-m}\D(\s)~\pa_\lam^m\D(\lam),~~~~~~~ ({\rm for}~~ \l\geq 1).
\eea
The Hamiltonian of this system is
\bea
H&=&\int d\lam~P(t,\lam)Q'(t,\lam)~-~\CL(t,0)
\nn\\
&=&\int d\lam~P(t,\lam)Q'(t,\lam)~+~\frac12~Q(t,0)~
(e^{\pa_\s^2}Q(t,\s))|_{\s=0}~-~
\frac{g}{3}~Q(t,0)^3.
\label{ph}
\eea
\vs

In the basis $(e_n(\lambda),e^n(\lambda))$, \bref{expbase},
the constraints and the Hamiltonian are
\bea
\psi^m&=&
-~\sum_{\l=0}^{\infty}~\frac{q^{m+2\l}}{\l!}~+~g~
\sum_{\l=0}^m { }_{m}C_\l~q^{m-\l}q^\l~\approx~ 0,
\label{peomexp}
\\
\varphi_{2m}&=&p_{2m}-\frac12~\sum_{\l=0}^{\infty}\frac{1}{(\l+m+1)!}~
q^{2\l+1}~
\approx~ 0,
\nn\\
\varphi_{2m+1}&=&p_{2m+1}+{\frac12}~\sum_{\l=0}^{\infty}
\frac{1}{(\l+m+1)!}~q^{2\l}~\approx~ 0,
\label{pmomexp}\\
H&=&\sum_{m=0}^{\infty}p_mq^{m+1}~+~\frac12~
q^0~\sum_{m=0}^{\infty}\frac{{q^{2m}}}{m!}~-~\frac{g}3(q^0)^3.
\label{phexp}
\eea
where ${ }_{{m}}C_\l=\frac{m!}{(m-\l)!\l!}~$ is the combinatorial coefficient.
Note that $\psi^m$ preserves its form in time.

\vs

\subsection{Free p-adic particle}
\indent

We first consider the case of free action.
When $g=0$
the \eom constraints \bref{peomexp} can be recombined as
\bea
\t\psi^{m}~\equiv~-~\sum_{r=0}^\infty\frac{(-)^r}{r!}~\psi^{2r+m}&=&{q^{m}}~
\approx~0.
\label{pfeomexp}
\eea
The momentum constraints \bref{pmomexp} are also simplified
by making combinations with the \eom constraints
\bea
\t\varphi_{2m}&\equiv&\varphi_{2m} ~-~\frac12\sum_{r=0}^\infty
\frac{(-)^r~\psi^{2 r+1}}
{m!r!(m+r+1)}~=~p_{2m}~\approx~ 0,
\nn\\
\t\varphi_{2m+1}&\equiv&\varphi_{2m+1}~+~\frac12\sum_{r=0}^\infty
\frac{(-)^r~\psi^{2 r}}
{m!r!(m+r+1)}~=~p_{2m+1}~\approx~ 0.
\label{pfmomexp}
\eea
The constraints $(\t\varphi_m,\t\psi^m)=(p_m,q^m)$
are second class. All canonical variables
$(q^m,p_m)$ vanish in the reduced phase space and therefore the
reduced phase space is 0-dimensional. 

{An analogous conclusion
    was obtained for a solution in the case of the purely cubic bosonic
    string field theory in \cite{Horowitz:1987kz}}.
The Hamiltonian of this system \bref{phexp}
vanishes in the physical phase space.
\vs


\subsection{Perturbative Pairing}
\noindent

We consider the reduced phase space around the fixed point $q=0$.
If we make the same combinations of constraints as
\bref{pfeomexp} and \bref{pfmomexp}
\bea
\t\psi^{m}&\equiv&-~\sum_{r=0}^{\infty}\frac{(-)^r}{r!}~\psi^{2r+m}
\nn\\&=&{q^{m}}~-~g~\sum_{r=0}^{\infty}\frac{(-)^r}{r!}
\sum_{\l=0}^{2r+m} { }_{{2r+m}}C_\l~q^{{2r+m}-\l}q^\l~\approx~ 0,
\label{ppeomexp}
\\
\t\varphi_{2m}&\equiv&\varphi_{2m}~-~
\frac12\sum_{s=0}^{\infty}(\frac{(-)^s}{(m+s+1)m!s!}~)\psi^{2s+1}
\nn\\&=&p_{2m}~-~
\frac{g}2~\sum_{s=0}^{\infty}(\frac{(-)^s}{(m+s+1)m!s!}~)
\sum_{\l=0}^{2s+1} { }_{{2s+1}}C_\l~q^{{2s+1}-\l}q^\l~\approx~ 0,
\nn\\
\t\varphi_{2m+1}&\equiv&\varphi_{2m+1}~+~
\frac12\sum_{s=0}^{\infty}(\frac{(-)^s}{(m+s+1)m!s!}~)\psi^{2s}
\nn\\&=&p_{2m+1}~+~
\frac{g}2~\sum_{s=0}^{\infty}(\frac{(-)^s}{(m+s+1)m!s!}~)
\sum_{\l=0}^{2s} { }_{{2s}}C_\l~q^{{2s}-\l}q^\l~\approx~ 0.
\label{ppmomexp}\eea
The only possible forms of the constraints
obtained by iterations are
\bea
\8\psi^m&=&q^m~\approx~0,~~~~~\8\varphi^m~=~p_m~\approx~0.
\eea
Therefore
the system has zero degrees of freedom and zero \ham as
in the free p-adic particle.
Obviously the symplectic two form \bref{twoform1} vanishes.
It means that at the local minimum $q=0$ there is no non-trivial solution.
That is there is no excitation modes at the local minimum (tachyonic vacuum).
Following Sen we may interpret this result saying that there are not 
closed string states at the minimum \cite{Sen:1999mh}.
\vs


\subsection{Physical space around the fixed point $q=\frac 1g$}
\noindent
The \eom constraints
\bref{ppeomexp} have non-perturbative fixed point  as well.
In order to see it we return to \bref{peom}
and note that it has a constant solution $q=\frac 1g$ in addition to the $q=0$
solution. We
introduce new canonical variables as
\bea
q^0&=&\frac{1}{g}~+~\t q^0,~~~~~q^j~=~\t q^j,~~~(j>0)
\nn\\
p^{2m}&=&\t p^{2m},~~~~~p_{2m+1}~=~\t p_{2m+1}-\frac{1}{2g(m+1)!}.
\label{pseparate}
\eea
In terms of new variables the constraints are
\bea
\t\psi^m&=&
2\t q^m~-~\sum_{\l=0}^\infty~\frac{\t q^{m+2\l}}{\l!}~+~g~
\sum_{\l=0}^{m} { }_{m}C_\l~\t q^{m-\l}\t q^\l~\approx~ 0,
\label{npeomexp}
\\
\t\varphi_{2m}&=&\t p_{2m}-\frac12~\sum_{\l=0}^\infty\frac{1}{(\l+m+1)!}~
\t q^{2\l+1}~\approx~ 0,
\nn\\
\t\varphi_{2m+1}&=&\t p_{2m+1}
+{\frac12}~\sum_{\l=0}^\infty\frac{1}{(\l+m+1)!}~\t q^{2\l}~\approx~ 0
\label{npmomexp}
\eea
Note that $ \frac d{dt}\t\psi^m=\t\psi^{m+1}$.
The Hamiltonian \bref{phexp} is
\bea
H&=&
\sum_{m=0}^{\infty}~\t p_{m}\t q^{m+1}+\frac1{6g^2}
-(\t q^0)^2-\frac{g}{3}( \t q^0)^3
+\frac12~\t q^0\sum_{m=0}^{\infty}\frac{{\t q^{2m}}}{m!}.
\label{nphexp}
\eea

As we have separated  the $g^{-1}$ term, $\t q^0$ and $\t q^j$'s are
expected to be regular in $g\to 0$ and they can  be
determined iteratively.
In matrix form the \eom constraints \bref{npeomexp} are
\bea
\left( \begin{array}{c} \t\psi^{0} \\ \t\psi^{2} \\ \t\psi^{4} \\ \t\psi^6 \\
\cdots \end{array} \right) &=& \pmatrix{1&-\frac{1}{1!}&-\frac{1}{2!}&-
\frac{1}{3!}&-\frac{1}{4!}&...\cr
0& 1&-\frac{1}{1!}&-\frac{1}{2!}&-\frac{1}{3!}&-\frac{1}{4!}&...\cr
0&0& 1&-\frac{1}{1!}&-\frac{1}{2!}&-\frac{1}{3!}&...\cr
0&0&0& 1&-\frac{1}{1!}&-\frac{1}{2!}&...\cr.......
} \left( \begin{array}{c} \t q^{0} \\ \t q^{2} \\ \t q^{4} \\ \t q^6 \\ \cdots \end{array} \right) + ~g~ \left( \begin{array}{c} \t F^{0} \\ \t F^{2} \\ \t F^{4} \\ \t F^6 \\ \cdots \end{array} \right) ~\approx~ 0,
\nn\\
\label{npeomexp20}\\
\left( \begin{array}{c} \t\psi^{1} \\ \t\psi^{3} \\ \t\psi^{5} \\ \t\psi^7 \\ \cdots \end{array} \right) &=& \pmatrix{1&-\frac{1}{1!}&-\frac{1}{2!}&-\frac{1}{3!}&-\frac{1}{4!}&...\cr
0& 1&-\frac{1}{1!}&-\frac{1}{2!}&-\frac{1}{3!}&-\frac{1}{4!}&...\cr
0&0& 1&-\frac{1}{1!}&-\frac{1}{2!}&-\frac{1}{3!}&...\cr
0&0&0& 1&-\frac{1}{1!}&-\frac{1}{2!}&...\cr.......
} \left( \begin{array}{c} \t q^{1} \\ \t q^{3} \\ \t q^{5} \\ \t q^7 \\ \cdots \end{array} \right) \nonumber + ~g~ \left( \begin{array}{c} \t F^{1} \\ \t F^{3} \\ \t F^{5} \\ \t F^7 \\ \cdots \end{array} \right) ~\approx~ 0,
\nn\\
\label{npeomexp21}
\eea
where
\bea
\t F^{l}=\sum_{m=0}^{l} {}_{l}C_m~\t q^{l-m}\t q^m~
\eea

In this case to find the pairing among the $\psi$ constraints and
$\varphi$ constraints is not obvious. Here we are going to use another
way that consists in solving the $\psi$ constraints
\footnote{ To see the procedure
in the simple case of the harmonic oscillator see the appendix A}.

Lets us indicate by $\CD$ the
matrix appearing in the equation \bref{npeomexp20} and \bref{npeomexp21}.
The matrix $\CD$ has unit determinant\footnote{We have assumed
a regularization of $\CD=\lim_{n\to\infty}\CD_n$.}
 and has a "formal inverse",
\bea
\CD^{-1}&=&
\pmatrix{1& \frac{1}{1!}&\frac{3}{2!}&\frac{13}{6}&...\cr
0&1& \frac{1}{1!}&\frac{3}{2!}&\frac{13}{6}&...\cr
0&0&1& \frac{1}{1!}&\frac{3}{2!}&...\cr
0&0&0&1& \frac{1}{1!}&...\cr .......
}=
\pmatrix{c_0&c_1&c_2&c_3&...\cr
0&c_0& c_1&c_2&c_3&...\cr
0&0&c_0& c_1&c_2&...\cr
0&0&0&c_0& c_1&...\cr .......
}
\eea
where $c_n$'s are generated by
\bea
\frac{1}{2-e^x}&=&\sum_{n=0}c_n~x^n~=~1+x+\frac32x^2+\frac{13}{6}x^3+....
\eea
The infinite matrix $\CD$ has an infinite number of eigen-vectors. In fact we
have
\bea
 \CD\pmatrix{1\cr \kappa^2 \cr  \kappa^4 \cr  \kappa^6 \cr  \kappa^8 \cr ...}
&=&(2-e^{\kappa^2})
\pmatrix{1\cr \kappa^2 \cr  \kappa^4 \cr  \kappa^6 \cr  \kappa^8 \cr ...},
\label{null}
\eea
and the eigen-value can vanish when $\kappa$ satisfies
\be
2~-~e^{\kappa^2}~=~0.
\label{npdisp}
\ee
This equation  has an infinite numbers of solutions
\bea
\kappa^2_r&=&\log 2+2\pi i r,~~~~~r\in Z.
\label{null3}
\eea

Using the null vectors \bref{null} with \bref{null3} we can rewrite the
constraints
\bref{npeomexp20} in the following equivalent forms
\bea
\left( \begin{array}{c} \t q^{0} \\ \t q^{2} \\ \t q^{4} \\ \t q^6 \\
\cdots \end{array} \right)&\approx &\sum_r \left( \begin{array}{c}
1 \\ \kappa_r^{2} \\ \kappa_r^{4} \\
\kappa_r^6 \\ \cdots \end{array} \right)~A^r(t)~-~g~
\pmatrix{1& \frac{1}{1!}&\frac{3}{2!}&\frac{13}{6}&...\cr
0&1& \frac{1}{1!}&\frac{3}{2!}&\frac{13}{6}&...\cr
0&0&1& \frac{1}{1!}&\frac{3}{2!}&...\cr
0&0&0&1& \frac{1}{1!}&...\cr .......
} \left( \begin{array}{c} \t F^{0} \\ \t F^{2} \\ \t F^{4} \\ \t F^6 \\ \cdots \end{array} \right),
\nn \\ \label{newel}
\eea
\bea
\left( \begin{array}{c} \t q^{1} \\ \t q^{3} \\ \t q^{5} \\ \t q^7 \\ \cdots \end{array} \right)&\approx &\sum_r \left( \begin{array}{c} \kappa_r^{1} \\ \kappa_r^{3} \\ \kappa_r^{5} \\ \kappa_r^7 \\ \cdots \end{array} \right)~A^r(t)~-~g~
\pmatrix{1& \frac{1}{1!}&\frac{3}{2!}&\frac{13}{6}&...\cr
0&1& \frac{1}{1!}&\frac{3}{2!}&\frac{13}{6}&...\cr
0&0&1& \frac{1}{1!}&\frac{3}{2!}&...\cr
0&0&0&1& \frac{1}{1!}&...\cr .......
} \left( \begin{array}{c} \t F^{1} \\ \t F^{3} \\ \t F^{5} \\ \t F^7 \\ \cdots \end{array} \right)
\nn \\ \label{eomconam2}
\eea
where the functions $A^r(t)$, appearing as coefficients of the null vectors,
should satisfy
\bea
\dot A^r(t)=\kappa_r~A^r(t)
\label{Adotpadic}
\eea
in order the constraints preserve their forms, \bref{newel} and
\bref{eomconam2}, in time.

By iterations in $g$,~  $\t q^j$ can be obtained in terms of the
 $A^r(t)$\footnote{Special care must be paid in $g^2$ term when two $\kappa$'s cancel in $\kappa_j+\kappa_k+\kappa_\l$ since the denominator $(e^{(\kappa_j+\kappa_k+\kappa_\l)^2}-2)$ vanishes in those cases. },
\bea
\t q^m&\approx&\sum_j A^{j}{(\kappa_j)^m}~+~
g\sum_{jk} \frac{A^{j} A^{k}}{(e^{(\kappa_j+\kappa_k)^2}-2)}
{(\kappa_j+\kappa_k)^m}~+~
\nn\\&+&g^2\sum_{jk\l} \frac{2A^{\l}A^{j} A^{k}}
{(e^{(\kappa_j+\kappa_k+\kappa_\l)^2}-2)(e^{(\kappa_j+\kappa_k)^2}-2)}
{(\kappa_\l+\kappa_j+\kappa_k)^m}~
+~....
\nn\\
&\equiv&f^m[A(t)].
\label{npsol}
\eea
It is the solution of \eom constraints \bref{npeomexp}
and gives the solutions of the EL equation
\bref{paELeq}.
The Moeller-Zwiebach solution corresponds to $A^0={\rm real},
\kappa_0=\sqrt{\log 2}$ and other $A^r=0$
\cite{Moeller:2002vx},
\bea
q(t) &=& {1 \over g} + A^0 e^{\kappa_0 t} + g {(A^0)^2 \over
 (e^{(2\kappa_0)^2} - 2)} e^{(2\kappa_0) t} + \nonumber \\
&+& g^2 {2(A^0)^3 \over (e^{(3\kappa_0)^2} - 2)(e^{(2\kappa_0)^2} - 2)}
 e^{(3\kappa_0)t} + \cdots
\nn\\
&=& {1 \over g} + A^0 e^{\kappa_0 t} + g {(A^0)^2 \over 14} e^{(2 \kappa_0) t}
+ g^2 {2 (A^0)^3 \over 510·14} e^{(3\kappa_0) t} + \cdots.
\label{solMZ}
\eea
Solutions with infinite number of parameters have also been
found by Schnabl, Sen and Zwiebach \cite{ssz}\footnote{We
are grateful to Martin Schnabl for discussions.}.
A solution with $A^1$, $A^{1*} \neq 0$ is given by
\bea
q(t) &=& {1 \over g} + \left(A^1 e^{(b_1 + ic_1)t} +
A^{1*} e^{(b_1 - ic_1)t}
\right) +
\nn\\
&+&g\left({(A^1)^2 \over (e^{4(b_1^2-c_1^2)} - 2)} e^{2(b_1+ic_1)t} +
{2 A^1 A^{1*} \over (e^{4 b_1^2} - 2)} e^{2b_1 t} +
{(A^{1*})^2 \over (e^{4(b_1^2-c_1^2)} - 2)} e^{2(b_1-ic_1)t}\right)+\cdots,
\nn\\
\label{solSMZ}\eea
where ${b_1^2-c_1^2} = \log 2$, $b_1c_1 = \pi N, (b_1>0)$ for some non-zero
integer $N$. The sums are taken over $A^j = A^1, A^{1*}$ and
$\kappa_1 = (b_1 + ic_1), \kappa_1^* = (b_1-ic_1)$.

\vskip 4mm

The momentum constraints \bref{npmomexp} can be also solved using \bref{npsol}
\bea
\t p_{2m}-\frac12~\sum_{\l=0}\frac{1}{(\l+m+1)!}~
f^{2\l+1}[A]~&\approx~&0,
\nn\\
\t p_{2m+1} 
+{\frac12}~\sum_{\l=0}\frac{1}{(\l+m+1)!}~f^{2\l}[A]~&\approx~&0.
\label{npmomexp2}
\eea
Since we have the solutions of the coordinates $q^m$ and the momenta
$p_m$ in terms of infinite arbitrary constants $A^r$
we could compute the symplectic two form \bref{twoform1} in terms of these
constants.
At lowest order in $g$
\bea
\W&=&\sum_m~dq^m\wedge~dp_m
\nn\\&=&\frac12~
\sum_{rs}dA^r\wedge dA^s~(\kappa_r-\kappa_s)~
[\sum_{m,\l=0}\frac{1}{(m+\l+1)!}
(\kappa_r)^{2\l}(\kappa_s)^{2m}].
\label{Wpadic0}
\eea
In the last sum terms with ${\kappa_s^2}\neq{\kappa_r^2}$ are
$\frac{e^{\kappa_r^2}-e^{\kappa_s^2}}{{\kappa_r^2}-{\kappa_s^2}}$ and
vanish using \bref{npdisp}. On the other hand
terms with ${\kappa_s^2}={\kappa_r^2}$ becomes $e^{\kappa_r^2}$. Then
$\W$ becomes
\bea
\W&=&
\sum_{rs}dA^r\wedge dA^s~(\kappa_r-\kappa_s)~|_{\kappa_s^2=\kappa_r^2}~
~=~\sum_{r}~2\kappa_r~dA^r\wedge dA^{-r}
\label{Wpadic}\eea
where only $\kappa_s+\kappa_r=0$ terms remain non-vanishing.

Similarly if we use the same formula in the Hamiltonian it becomes ,
at lowest order in $g$ ,
\bea
H&=&\frac{1}{6g^2}~-~
\frac12\sum_{rs}A^rA^s~\kappa_r(\kappa_r-\kappa_s)~
[\sum_{m,\l=0}\frac{1}{(m+\l+1)!}
(\kappa_r)^{2\l}(\kappa_s)^{2m}]
\nn\\&=&
\frac{1}{6g^2}~-~2~\sum_r~\kappa_r^2~A^rA^{-r}.
\label{hhoexp0}
\eea
\vs

The forms of the symplectic two form $\W$ \bref{Wpadic} and the Hamiltonian
$H$ \bref{hhoexp0} show that the p-adic particle has an infinite dimensional
physical phase space around the fixed point $\frac 1g$.

So far we are assuming $q(t)$ to be any sufficiently smooth functions.
In the discussions of tachyon condensation we are interested in the field
$q(t)$ which start from the local maximum $q(t)=\frac1g$ at $t=-\infty$.
It restricts the modes only with
\bea
Re(\kappa_r)~>~ 0.
\label{BC}
\eea
The solutions \bref{solMZ} and \bref{solSMZ} are satisfying this requirement.
In restricting the function space by imposing \bref{BC}
there appears no term satisfying $\kappa_s+\kappa_r=0$ in the sum in
\bref{Wpadic0}. The symplectic two form $\W$ \bref{Wpadic} vanishes
identically. Similarly the  Hamiltonian
$H$ \bref{hhoexp0} becomes a constant $\frac1{6g^2}$, the asymptotic value of
energy at  $t=-\infty$. (They are shown explicitly up to $g^2$.)

These results imply that
the unstable and stable
manifold of the non-perturbative fixed point
$q=\frac 1g$ is a infinite lagrangian submanifold of $T^*J$.
The manifold of solutions
relevant for the rolling tachyon, unstable submanifold, is infinite
dimensional without a phase space structure. 
See fig. 1 for a summary of these results.

If we compute the energy \bref{hhoexp0} without imposing any
  boundary conditions, solutions do not have a definite sign of the
  energy as we will see in Appendix B. 
It corresponds to the general property of the non-local
  theories.
\vs

\section{String Field theory}
\noindent

In order to study the rolling of tachyon we need to know
the potential. The natural framework to compute it in string theory is
the string field theory. The level truncation is an approximation  scheme
\cite{Kostelecky:1989nt} to compute the potential.
For the lowest level truncation approximation
and  for spatially homogeneous configurations the action is given by
\be
S=
\int dt ({\frac12}\dot q(t)^2 +\frac{\M^2}{2}q(t)^2
-{g\over 3} {\tilde q}(t)^3 )
\ee
where
\bea
\tilde{q}(t) = e^{-\partial^2_t} q (t).
\eea
The Euler-Lagrange equation is
\bea
\ddot q(t)~ {-~\M^2} q(t)~+~g~e^{-\partial^2_t}(e^{-\partial^2_t}q(t))^2~=~0.
\label{seomq}
\eea
{It has been examined as a non-local system in \cite{Eliezer:1989cr}.}

In the present formalism the associated \lag density $\CL(t,\lam)$ is
\bea
 \mathcal{L}(t,\lam) &=& {\frac12}Q'(t,\lam)^2+\frac{\M^2}2 Q(t,\lam)^2
-{g\over 3} \tilde Q(t,\lam)^3
\eea
with
\bea
\tilde{Q}(t,\lam) = e^{-\partial^2_\lam}Q(t,\lam).
\nn
\eea

The \eom constraint \bref{eom} is
\bea
\psi(t,\lam)&=&
{-}\pa_{\lam}^2Q(t,\lam)~ {+~\M^2} Q(t,\lam)~-~g~e^{-\partial^2_\lam}
(\tilde{Q}(t,\lam))^2~\approx~0,
\label{seom}\eea
the momentum constraint \bref{mom} is
\bea
\varphi(t,\lam) &=&
P(t,\lam) - Q'(t,0) \D(\lam)  ~ {+~g~} \int d\s
\sum_{n=0}^{\infty} \sum_{m=0}^{\infty} {(-1)^{n+m+1}\over (n+m+1)!}
\nn\\&&
\bigg[\pa^{2n+1}_{\s} \D(\s) (-\pa_{\lam})^{2m}\D(\lam)
+ \pa^{2n}_{\s} \D(\s) (-\pa_{\lam})^{2m+1} \D(\lam)
\bigg] (\tilde Q(t,\s))^2 ~\approx~0
\nn
\label{smom}
\eea
and the Hamiltonian \bref{h} is
\be
H=\int d\lam P(t,\lam)Q^\prime(t,\lam)-
{\frac12}Q'(t,0)^2-\frac{\M^2}2 Q(t,0)^2
+{g\over 3} \t Q(t,0)^3.
\label{sh}
\ee

In the basis \bref{QPexp} the constraints are given by
\bea
\varphi_{2m}&=&p_{2m}~-~q^1~\D_{m,0}~-~
2~g~\sum_{r=0}^{\infty}~\sum_{s=0}^{\infty}~
B_{m,r,s}^{eo}~q^{2r}~q^{2s+1},
\label{smomexpev}
\nn\\
\varphi_{2m+1}&=&p_{2m+1}~+~g~\sum_{r=0}^{\infty}\sum_{s=0}^{\infty}(
B_{m,r,s}^{ee}~q^{2r}~q^{2s}+
B_{m,r,s}^{oo}~q^{2r+1}~q^{2s+1}),
\label{smomexpod}
\\
\nn\\
\psi^{2\l}&=&-q^{2\l+2}~+~\M^2q^{2\l}~+~
\nn\\&&~~~~~~~~-g~\sum_{r=0}^{\infty}\sum_{s=0}^{\infty}(B_{-\l-1,r,s}^{ee}
q^{2r}~q^{2s}+
B_{-\l-1,r,s}^{oo}q^{2r+1}~q^{2s+1}),
\label{seomexpev}
\nn\\
\psi^{2\l+1}&=&-q^{2\l+3}~+~\M^2q^{2\l+1}~-~
2g~\sum_{r=0}^{\infty}~\sum_{s=0}^{\infty}~B_{-\l-1,r,s}^{eo}~
q^{2r}~q^{2s+1},
\label{seomexpod}
\eea
where
\bea
B_{m,r,s}^{eo}&\equiv&
\sum_{t=0}^{r}~\sum_{n=0}^{s}~
\frac{(-1)^{m+r+s+1}}{\Gamma(m+n+t+2)(r-t)!(s-n)!}{~}_{2n+2t+1}C_{2t},
\label{Beo}
\nn\\
B_{m,r,s}^{ee}&\equiv&
\sum_{t=0}^{r}~\sum_{n=0}^{s}~
\frac{(-1)^{m+r+s+1}}{\Gamma(m+n+t+2)(r-t)!(s-n)!}{~}_{2n+2t}C_{2t},
\label{Bee}
\nn\\
B_{m,r,s}^{oo}&\equiv&
\sum_{t=0}^{r}~\sum_{n=0}^{s}~
\frac{(-1)^{m+r+s}}{\Gamma(m+n+t+3)(r-t)!(s-n)!}{~}_{2n+2t+2}C_{2t+1}.
\label{Boo}
\eea
The Hamiltonian \bref{sh} is
\bea
H&=&\sum_{m=0}^\infty~p_mq^{m+1}~-~\left(
\frac12 (q^1)^2+\frac{\M^2}2 (q^0)^2-\frac{g}{3}
(\sum_{n=0}^\infty\frac{(-1)^nq^{2n}}{n!})^3\right).
\label{shexp}
\eea
\vs


\subsection{Perturbative case}
\indent

When the coupling constant $g$ is zero , the system becomes
that of the simple harmonic oscillator, with negative $\w^2$ ($\w^2=-\M^2$),
discussed in the subsection 3.2.
The phase space around the fixed point $q=0$ is
described by $(q^0,p_0)$ and is
two dimensional.

If we use the  perturbative expansion
all the constraints remain in the second class.
Then the degrees of freedom will not change and the dimension is equal
to $2$ \cite{Eliezer:1989cr}\cite{lv}.
\vs

$q=0$ is unstable local maximum. The above 2 degrees of freedom are
unstable modes
\bea
q(t)&=&a^{1}e^{Mt}+a^{-1}e^{-Mt}+O(g).
\eea
The \eom constraints are used iteratively to find solutions in terms of
two constants $a^{\pm 1}.$ \footnote{A numerical analysis of time
dependent solutions of string field
theory with one arbitrary function was given by Fujita and Hata
\cite{fujitahata:2003fh}.}
The symplectic two form and the Hamiltonian for this case are
\bea
\W&=&2M~da^1\wedge da^{-1}~+~O(g),
\eea
\bea
H=&=&-2M^2~a^1a^{-1}~+~O(g).
\eea
The $a^{1}$ mode describes rolling solution satisfying $q(-\infty)=0$.
Under this boundary condition
$Q(\lam)$ is given in a similar form as  in the p-adic particle
\cite{Moeller:2002vx}
\bea
q(t)=\sum_{n=1}^\infty a^n(a^1) e^{nMt}
\eea
where $a^n$ is determined in terms of $a^1$ by the recursive relation
\bea
a_n&=&\frac{g}{(1-n^2)M^2}\sum_{j=1}^{n-1}e^{-2(n^2-nj+j^2)M^2}~a_ja_{n-j},~~
(n\geq 2).
\eea
Note that the stable and unstable lagrangian submanifold of the
fixed point $q=0$
are both one dimensional.


\subsection{Physical space around $q=\frac{\M^2}{g}$}
\indent

We are interested in solving the non-linear constraint of
the tachyon particle \bref{seom}. It has two constant solutions,
$q=0$ and $q=\frac{\M^2}{g}$. The first one is corresponding
to the perturbative case of the last subsection. Here we examine
the perturbation around the second solution.

New canonical coordinates are introduced as
\bea
q^0&=&\frac{\M^2}{g}~+~\t q^0,~~~~~q^j~=~\t q^j,~~~(j>0)
\nn\\
p_{2m+1}&=&\t p_{2m+1}-{\frac{\M^4}{g}}\frac{(-1)^{m+1}}{(m+1)!},~~~~~
p_{2m}~=~\t p_{2m}.
\label{separate2}
\eea
In terms of these variables the constraints are
\bea
\psi^{2\l}&=&-\t q^{2\l+2}~+~\M^2\t q^{2\l}~-~2\M^2\sum_{s=0}^\infty
\frac{(-2)^{s}\t q^{2s+2\l}}{s!}
\nn\\&& \hskip 30mm -
g~\sum_{r=0}^{\infty}~\sum_{s=0}^{\infty}~(B_{-\l-1,r,s}^{ee}~
\t q^{2r}~\t q^{2s}
+B_{-\l-1,r,s}^{oo}~\t q^{2r+1}~\t q^{2s+1}),
\nn\\
\psi^{2\l+1}&=&-\t q^{2\l+3}+\M^2\t q^{2\l+1}-2\M^2\sum_{s=0}^\infty
\frac{(-2)^{s}\t q^{2s+2\l+1}}{s!}\\&& \hskip 30mm
-2g~\sum_{r=0}^{\infty}~\sum_{s=0}^{\infty}~B_{-\l-1,r,s}^{eo}~
\t q^{2r}~\t q^{2s+1}.
\nn\\
\label{nseomexp}
\eea
The momentum constraints  \bref{smomexpod} are
\bea
\varphi_{2m}&=&
\t p_{2m}~-~\t q^1~\D_{m,0}~-~2~\M^2~\sum_{s=0}^{\infty}~
B_{m,0,s}^{eo}~\t q^{2s+1}~\\&& \hskip 30mm
-~2~g~\sum_{r=0}^{\infty}~\sum_{s=0}^{\infty}~
B_{m,r,s}^{eo}~\t q^{2r}~\t q^{2s+1},
\label{momconevs}
\nn\\
\varphi_{2m+1}&=&\t p_{2m+1}~+~2\M^2~\sum_{s=0}^{\infty}~
B_{m,0,s}^{ee}~\t q^{2s}
\nn\\&&~~~~~~~~~~~~+~g~\sum_{r=0}^{\infty}~\sum_{s=0}^{\infty}~(
B_{m,r,s}^{ee}~\t q^{2r}~\t q^{2s}
+B_{m,r,s}^{oo}~\t q^{2r+1}~\t q^{2s+1}).
\label{nsmomexp}
\eea

For $g=0$ the \eom constraints \bref{nseomexp} have solutions
\bea
\t q^m&\approx&
\sum_r A^{r}~{(\kappa_r)^m}
\eea
where $A^r(t)$'s are arbitrary functions of time and $\kappa_r$'s are
infinite number of solutions of
\bea
\kappa^2-\M^2+2\M^2 e^{-2\kappa^2}~=~0.
\label{strdisp}
\eea
There are infinite number of complex quartet  solutions
\bea
\{~\kappa_N,~\kappa_N^*,~-\kappa_N,~-\kappa_N^*,~\},~~~
N=1,2,3,...
\eea
Two real solutions $\pm\kappa_0=\pm\sqrt{-\frac12{\rm
ProductLog}[-1,-\frac1{2e}]}$  are present only for a value of
$M^2=-\frac12(1+{\rm ProductLog}[-1,-\frac1{2e}])\approx 0.916.$
In a power series on $g$ the solutions of the \eom constraints are given by
\bea
\t q^m&\approx&
\sum_r A^{r}_{(0)}~{(\kappa_r)^m}~+~
g~\sum_{rs} A^{rs}_{(1)}~{(\kappa_r+\kappa_s)^m}~
+~g^2~
\sum_{rst} A^{rst}_{(2)}~(\kappa_r+\kappa_s+\kappa_t)^m~+~
\nn\\
&\equiv&f^m[A],
\label{gsolution3}
\eea
where
\bea
A^{r}_{(0)}&=&A^r,~~~~~
A^{rs}_{(1)}~=~ \frac{-A^{r} A^{s}
e^{-2(\kappa_r^2+\kappa_r\kappa_s+\kappa_s^2)}}
{(\kappa_r+\kappa_s)^2-\M^2+2\M^2e^{-2(\kappa_r+\kappa_s)^2}},
\nn \\
A^{rst}_{(2)}&=&
\frac{2~A^{r}A^s A^{t}e^{-(\kappa_r+\kappa_s+\kappa_t)^2}e^{-\kappa_r^2}
e^{-(\kappa_s+\kappa_t)^2}
e^{-2(\kappa_s^2+\kappa_s\kappa_t+\kappa_t^2)}}
{((\kappa_r+\kappa_s+\kappa_t)^2-\M^2+2\M^2
e^{-2(\kappa_r+\kappa_s+\kappa_t)^2})
((\kappa_s+\kappa_t)^2-\M^2+2\M^2e^{-2(\kappa_s+\kappa_t)^2})}
\nn \\ \label{gsolution32}
\eea
Eqs. \bref{gsolution3} and \bref{gsolution32} give the solution of the EL
equation \bref{seomq}.

The \eom constraints can be solved as
\bea
&&\t q^m-f^m[A]\approx 0,~~~~~~~~~(m\geq 0),
\label{nseomexp2}
\eea
where
the functions $A^r(t)$  satisfy
\bea
\dot A^r(t)=\kappa_r~A^r(t)
\label{padota}
\eea
as in the p-adic case.

The  momentum constraints \bref{momconevs} are solved

using \bref{nseomexp2}
\bea
\t p_{2m}&-&f^1[A]~\D_{m,0}~-~2~\M^2~\sum_{s=0}^{\infty}~
B_{m,0,s}^{eo}~f^{2s+1}[A]~
\nn\\&-&
2~g~\sum_{r=0}^{\infty}~\sum_{s=0}^{\infty}~
B_{m,r,s}^{eo}~f^{2r}[A]~f^{2s+1}[A]\approx 0,
\label{momconevs2}
\nn\\
\t p_{2m+1}&+&2\M^2~\sum_{s=0}^{\infty}~
B_{m,0,s}^{ee}~f^{2s}[A]
\nn\\&+&g~\sum_{r=0}^{\infty}~\sum_{s=0}^{\infty}~(
B_{m,r,s}^{ee}~f^{2r}[A]~f^{2s}[A]
+B_{m,r,s}^{oo}~f^{2r+1}[A]~f^{2s+1}[A])\approx 0.
\nn\\ \label{nsmomexp2}
\eea

Since we have the solutions of the coordinates $q^m$ and  momenta
$p_m$ in terms of infinite arbitrary constants $A^r$
we can compute the symplectic two form \bref{twoform1} in terms of these
constants.
The computation is similar to the p-adic particle case and
at lowest order in $g$
\bea
\W&=&\sum_m~dq^m\wedge~dp_m
\nn\\&=&
\sum_{r}~\kappa_r~(1-4M^2e^{-2\kappa_r^2})~dA^r\wedge dA^{-r}~.
\label{Wstring0}
\eea
Similarly the Hamiltonian in \bref{shexp}
\bea
H&=&
\sum_{j=0}^\infty~\t p_j\t q^{j+1}~-~{\frac{\M^6}{6g^2}}~-~
\frac12 (\t q^1)^2~-~\frac{\M^2}2 (\t q^0)^2~
\nn\\&&+~
\M^2(\sum_{n=0}^\infty\frac{(-1)^n \t q^{2n}}{n!})^2~+~\frac{g}{3}
(\sum_{n=0}^\infty\frac{(-1)^n \t q^{2n}}{n!})^3
\label{nshexp}
\eea
is, at lowest order in $g$ ,
\bea
H&=&
-\frac{M^6}{6g^2}~-~\sum_r~(1-4M^2e^{-2\kappa_r^2})~\kappa_r^2~A^rA^{-r}.
\label{hsft0}
\eea
The forms of the symplectic two form $\W$ \bref{Wstring0} and the Hamiltonian
$H$ \bref{hsft0} show that this is an infinite dimensional system.
This implies
that there are infinite continuous excitations around the tachyon vacuum.

So far we are assuming $q(t)$ to be any sufficiently smooth functions.
If  we are interested in the field
$q(t)$ which approaches to the tachyonic vacuum at $t=+\infty$.
It restricts the modes with
\bea
Re(\kappa_r)~<~ 0.
\label{BCsft}
\eea
In restricting the function space by imposing \bref{BCsft}
there appears no term satisfying $\kappa_s+\kappa_r=0$ in the sum in
\bref{Wstring0}. The symplectic two form $\W$ \bref{Wstring0} vanishes
identically. Similarly the  Hamiltonian
$H$ \bref{hsft0} becomes a constant $-\frac{M^6}{6g^2}$, asymptotic value of
energy at  $t=+\infty$. (They are shown to be valid up to $g^2$.)
They show that the system with the boundary condition 
is an infinite dimensional lagrangian submanifold. 
Following Sen's conjecture \cite{Sen:1999mh} these excitations around the
closed string vacuum may be interpreted as closed string states. This
interpretation is not clear among other reasons because the possible 
closed string states would correspond to the quantization of the above 
solutions and we have not perform this quantization. 
\vs

The rolling solution is $q=0$ at $t=-\infty$ and
it is expected to pass through $q=\frac{M^2}{g}$ at $t=+\infty$.
The initial energy should be  $H=0$, which is different from the above 
non-perturbative one with energy $H=-\frac{M^6}{6g^2}$.} 
Since we have a conservation of energy
within our model this is not possible, it means that the tachyon will not
 stop at the minimum but it will oscillate as in the p-adic case.
 See fig. 2 for a summary of string field theory results.

As in the p-adic case, if we evaluate the energy without imposing any
boundary condition the solutions do not have a definite sign of the energy
as is seen in Appendix C.
\vs


\section{Discussions}
\indent

The 1+1 dimensional Hamiltonian formulation of non-local theories is a
valuable tool to construct the 
physical reduced phase space of a non-local theory.
The idea is similar to the construction of the physical phase space of
gauge theories.

We have shown that there are two possible physical phase spaces, perturbative
and non-perturbative around fixed points,
these dimensions do not coincide in general.
In particular for the case of the p-adic particle around the fixed
point $q=0$
the physical space is
zero dimensional at perturbative level. 
There are no excitation modes around this
tachyon vacuum.
For the fixed point $q=\frac 1g$
the perturbative phase space is infinite dimensional. The
solutions contain an infinite number of constants with a non-vanishing
symplectic structure.
For the rolling type solutions the physical space is infinite 
dimensional lagrangian submanifold instead.

In the case of string field
theory at lowest truncation level the perturbative phase space around $q=0$
is two dimensional, which coincides with the dimension of the free theory, 
and for the non-trivial fixed point $q=\frac {M^2}g$ it
is perturbatively infinite dimensional. 
There are continuous excitations around the tachyon vacuum.

The manifold of the solutions for the
p-adic particle 
and string field theory case is different. 
Around local maxima the space of solutions of the p-adic particle is
infinite dimensional while string field is two dimensional. 
Around local minima (tachyonic vacua) the physical excitations for 
the p-adic particle is
zero dimensional while string field is infinite dimensional. 
The different phase space structure is due to the 
dispersion relation for the solutions we found, see eq (83) and  eq (116).

It is also noted that in both cases, p-adic and string field theory, 
the solutions obtained without imposing any boundary condition 
have infinite dimensional physical phase space and do not have a
definite sign for the energy, which in addition, is not  bounded from below.

It would be interesting to examine solutions describing the 
tachyon condensation in these models
by introducing dissipation of energy suitably.

\vskip 10mm

{\bf Acknowledgements}

We are grateful to Roberto Casalbuoni, Alex Haro, Jaume Gomis,
Shamit Kachru, Josep Llosa, Toni  Mateos, Hirosi Ooguri, Josep
Maria Pons, Jorge Russo, Martin Schnabl, Paul Townsend and Peter
West, Richrad Woodard. 
We acknowledge the Benasque Center of Science where a part
of this work has been done. This work is supported in part by the
European Community's Human Potential Program under contract
HPRN-CT-2000-00131 Quantum Spacetime.
 This work is also supported by MCYT
FPA, 2001-3598 and CIRIT GC 2001SGR-00065.

\vs

\appendix

\section{Harmonic oscillator}

The Hamiltonian and constraints of  simple harmonic oscillator are,
\bref{simple}
\bea
H&=&p_mq^{m+1}~-~\frac12(q^1)^2~+~\frac{w^2}{2}(q^0)^2
\label{hho}\\
\varphi_0&=&p_0-q^1~\approx~0,~~~~~~~~~\varphi_j~=~p_j~\approx~0,~~~(j>0),
\\
\psi^m&=&q^{m+2}~+~w^2~q^m~\approx~0,~~~(m\geq 0).
\label{HOeom}
\eea

\eom constraint \bref{HOeom} is expressed as
\bea
\pmatrix{\psi^0\cr \psi^1\cr \psi^2\cr \psi^3\cr..}
&=&\pmatrix{\w^2&0&1&...\cr 0&\w^2&0&1&...\cr0&0& \w^2&0&1&...\cr
0&0&0& \w^2&0&1&...\cr...}
\pmatrix{q^0\cr q^1\cr q^2\cr q^3\cr..}
\eea
The coefficient matrix ${\cal D}$ has two zero vectors
\bea
\pmatrix{1\cr (iw)\cr (iw)^2\cr (iw)^3\cr..},~~~~~~~~~~
\pmatrix{1\cr (-iw)\cr (-iw)^2\cr (-iw)^3\cr..}
\eea
Then the general solution of the linear constraints \bref{HOeom} is
\bea
q^m(t)~-~A(t){(iw)^m}+A^*(t){(-iw)^m}\approx 0.
\label{HOsol}\eea
where the functions $A(t)$, appearing as coefficients of the null vectors,
should satisfy
\bea
\dot A(t)=i\w A(t)
\label{Adotpadic2}
\eea
in order \bref{HOsol} is consistent with
the Hamilton's equation $\frac{ d}{dt}q^m=q^{m+1}$.

The momentum constraints are solved by
\noindent
\bea
&&p_0 -(A(t){(iw)}+A^*(t){(-iw)}) \approx 0~~~~
\nn\\
&&p_n ~\approx~0,~(n>0)
\label{npmomexp21}\eea
All phase space variables $(q^m,p_m)$ are described in terms of $(A,A^*)$.

Note the Hamiltonian \bref{hho}, using the expressions
\bref{HOsol} and \bref{npmomexp21}, becomes
\bea
H&=&2\w^2 A^*A.
\label{hamhoexp0}
\eea
The symplectic two form $\W$ of this system is
\bea
\W&=&\sum_{m=0}dp_m\wedge dq^m~=~2i\w~dA^*\wedge dA.
\label{hamomega}
\eea
They show  $(A,A^*)$ plays a role of the canonical pair with dimension two.

\vs

\section{Energy of the p-adic particle
}

We have shown the p-adic particle around the non-perturbative fixed point 
$g=\frac1{g}$ is described by the Hamiltonian \bref{hhoexp0} 
\bea
H&=&
\frac{1}{6g^2}~-~2~\sum_r~\kappa_r^2~A^rA^{-r}
\label{hhoexp0a}
\eea
with the canonical form \bref{Wpadic}
\bea
\W&=&\sum_{r}~2\kappa_r~dA^r\wedge dA^{-r},
\label{Wpadica}\eea
where sums are taken over all modes of solutions of \bref{npdisp},
$e^{\kappa^2}-2=0$. 
There are infinite number of complex quartet  solutions
\bea
\{~\kappa_N,~\ba{\kappa_N},~-\kappa_N,~-\ba{\kappa_N},~\},~~~
N=0,1,2,3,...
\eea
where
$
\kappa_{N}~=~a_N+ib_N,~~~ 
a_N^2-b_N^2=\log~2,~~2a_Nb_N~=~2N\pi, \;\; a_N>0,~~b_N~\geq~0.
$ 
For $N=0$, $b_0=0$ then there are only two real solutions 
$\pm\kappa_0$.
We use these modes 
in the symplectic form $\W$ \bref{Wpadica} and the 
Hamiltonian \bref{hhoexp0a} explicitly they become
\bea
\W&=&4\sum_{N>0}~[\kappa_{N}~dA^{N}\wedge dA^{-N}+{\ba{\kappa_{N}}}~
d{\ba {A^{N}}}\wedge d{\ba{ A^{-N}}}]+4\kappa_{0}~dA^{0}\wedge dA^{-0}
\nn\\
H&=&{1\over6g^2}~-~4~\sum_{N>0}~\left[{\kappa_{N}}^2~A^{N}A^{-N}~+~
\overline{\kappa_{N}}^2~
  \overline{A^N}~ \overline{A^{-N}} \right]
-4\kappa_{0}^2~A^{0}A^{-0}.
\eea
Now, if we write the variables in terms of real coordinates 
as 
\be
A^{N}~=~E_N+iB_N,~~~~ A^{-N}~=~C_N+iD_N,~~~~
A^{0}~=~E_0,~~~~ A^{-0}~=~C_0,
\ee
the symplectic form $\W$ tells that 
the canonical momenta conjugate to $C$ and $D$
are
\be
P_{C_N}~=~8(a_NE_N-b_NB_N), \;\;\;\;\; P_{D_N}=8(-a_NB_N-b_NE_N),~~~~~
P_{C_0}~=~4a_0E_0.
\ee
Then we obtain
\bea 
\W&=&\sum_{N>0}~[dP_{C_N}\wedge dC_N+dP_{D_N}\wedge dD_N
]+dP_{C_0}\wedge dC_0,
\label{Wpadic3}
\\
\label{mopad}
H&=&{1 \over 6g^2}~-~\sum_{N>0}\left[(P_{C_N},P_{D_N})\left(\begin{array}{cc}
    a_N&-b_N\\b_N&a_N \end{array} \right) 
\left( \begin{array}{c} C_N\\D_N\end{array}
    \right)\right]-a_0P_{C_0}C_0.
\eea
Since the $C,D$ and their conjugate $P_C,P_D$ are real canonical variables 
the Hamiltonian is not bounded from below.
Using  symplectic transformations the Hamiltonian can further be simplified.

Using a symplectic transformation
\bea
C~=~{1\over \sqrt{2}}(q+p'), \;\;\;\;\; P_C~=~{1\over \sqrt{2}}(p-q'), \\
D~=~{1\over \sqrt{2}}(q'+p), \;\;\;\;\; P_D~=~{1\over \sqrt{2}}(p'-q),
\eea
the Hamiltonian \bref{mopad} becomes
\bea
H~=~{1\over6g^2}&+&{1\over2}\sum_{N>0}\left[(p,p')\left(\begin{array}{cc} b
    & -a \\ -a & -b \end{array} \right) \left(\begin{array}{c} p \\ p'
  \end{array} \right)~+~(q,q')\left(\begin{array}{cc} b
    & a \\ a &- b \end{array} \right) \left(\begin{array}{c} q \\ q'
  \end{array} \right)\right]_N
\nn\\&+&\frac{a}{2}(p_0^2-q_0^2).
\eea
After rotation of $(q,q')$, then rescaling the coordinates, it follows
\bea H={1 \over 6g^2}&+&{1 \over
2}\sum_{N>0}[(\tilde{p},\tilde{p}') \left(\begin{array}{cc} 1 & 0
\\ 0 & -1 \end{array}\right) \left(\begin{array}{c} \tilde{p}
\\ \tilde{p}' \end{array}\right) \nonumber \\ &+&
(\tilde{q},\tilde{q}') \left(\begin{array}{cc}
-(a^2-b^2) & 2ab \\ 2ab & (a^2-b^2)
\end{array}\right) \left(\begin{array}{c} \tilde{q} \\ \tilde{q}'
\end{array}\right)]_N~+~\frac{1}{2}(\tilde p_0^2- a_0^2 \tilde q_0^2).
\eea
The $p^2$ term and  ${p'}^2$ term have opposite signs.
There remains a freedom of $SO(1,1)$ rotation, 
which keeps kinetic terms invariant. However it can not eliminate the
$\tilde{q}\tilde{q}'$ term and the Hamiltonian is not diagonalized
completely.

Eigen frequencies are given from
\be {d^2 \over dt^2} \left( \begin{array}{c} \tilde{q} \\
\tilde{q}' \end{array} \right)~=~\left( \begin{array}{cc} (a^2-b^2) & -2ab \\
2ab & (a^2-b^2) \end{array} \right)\left( \begin{array}{c} \tilde{q} \\
\tilde{q}' \end{array} \right) \ee
The eigenvalues of the matrix are $(a_N\pm ib_N)^2=\kappa_N^2$  
in agreement with the solution \bref{Adotpadic} and \bref{npsol}.
Summing up, it can be seen that the non locality shows indefinite signs of
the energy. This agrees with the general discussion of the Ostrogradski
\cite{o}
in \cite{Eliezer:1989cr}.


\section{Energy for the String Field Theory
}

We can analyze  the string field theory case as in the p-adic particle.
Now we consider the 
local minimum $q={M^2 \over g}$ (tachyonic vacuum) rather than the
local maximum. 
The
symplectic form is \bref{Wstring0} 
\bea
\W&=&
\sum_{r}~\kappa_r~(1-4M^2e^{-2\kappa_r^2})~dA^r\wedge dA^{-r}~.
\label{Wstring0a}
\eea
and the Hamiltonian is \bref{hsft0} 
\bea
H&=&
-\frac{M^6}{6g^2}~-~\sum_r~(1-4M^2e^{-2\kappa_r^2})~\kappa_r^2~A^rA^{-r},
\label{hsft0a}
\eea
where 
$\kappa_r$'s are solutions of \bref{strdisp}, 
$\kappa^2-\M^2+2\M^2 e^{-2\kappa^2}=0$. It have an 
infinite number of complex quartet  solutions
\bea
\{~\kappa_N,~\ba{\kappa_N},~-\kappa_N,~-\ba{\kappa_N},~\},~~~
N=1,2,3,...
\eea
The symplectic form $\W$ \bref{Wstring0a} and the 
Hamiltonian \bref{hsft0a} are
\bea
\W&=&\sum_{N>0}~[\kappa_{N}(1-4M^2e^{-2\kappa_{N^2}})~dA^{N}\wedge dA^{-N}+
{\ba{\kappa_{N}(1-4M^2e^{-2\kappa_{N^2}})}}~
d{\ba {A^{N}}}\wedge d{\ba{ A^{-N}}}]
\nn\\
H&=&-{M^6\over6g^2}-2\sum_{N>0}\left[{\kappa_{N}}^2
(1-4M^2e^{-2\kappa_{N^2}})A^{N}A^{-N}+
\overline{\kappa_{N}^2(1-4M^2e^{-2\kappa_{N^2}})}~
  \overline{A^N}~ \overline{A^{-N}} \right].
\nn\\
\eea
Now we write them in terms of real coordinates as 
\be
A^{N}~=~E_N+iB_N,~~~~ A^{-N}~=~C_N+iD_N,~~~~
\ee
the symplectic form $\W$ tells that 
the canonical momenta conjugate to $C$ and $D$
are
\be
P_{C_N}~=~8(\tilde a_NE_N-\tilde b_NB_N), \;\;\;\;\; 
P_{D_N}=8(-\tilde a_NB_N-\tilde b_NE_N),
\ee
where
\be \kappa_{N}~=~a_N~+~ib_N, \;\;
\kappa^2_{N}(1-4M^2e^{-2\kappa^2_{N}})~
\equiv~2(\tilde{a}_N~+~i\tilde{b}_N).
\ee
Then we obtain
\bea 
\W&=&\sum_{N>0}~[dP_{C_N}\wedge dC_N+dP_{D_N}\wedge dD_N
],
\label{Wstring3}
\\
\label{mopads}
H&=&-{M^6\over 6g^2}~-~
\sum_{N>0}\left[(P_{C_N},P_{D_N})\left(\begin{array}{cc}
    a_N&-b_N\\b_N&a_N \end{array} \right) 
\left( \begin{array}{c} C_N\\D_N\end{array}
    \right)\right].
\eea
They have same forms as the p-adic particle case.
The $C,D$ and their conjugate $P_C,P_D$ are real canonical variables 
therefore the Hamiltonian is not bounded from below.


\end{document}